\documentclass[a4paper]{jpconf}
\usepackage{graphicx}

\usepackage{bm}
\usepackage{latexsym}
\usepackage{dcolumn}
\usepackage{amsmath,amsfonts,amssymb}
\usepackage{graphicx,epsfig}
\usepackage{verbatim}
\usepackage{color}
\usepackage[caption=false]{subfig}
\usepackage{slashed}
\usepackage{etex}
\usepackage{float}
\def\nn{\nonumber}

\def\l{\left}
\def\r{\right}
\def\DM{\mathrm{d}}
\def\lp{\ell_0}

\def \lp {L_0}

\def \lp {\ell_0}

\def \Rsq {\widetilde {\rm \bf Ric}(p;p_0)}
\def \Rsqn {\widetilde {\rm \bf Ric}}
\def \Rsg {{\rm \bf Ric}(p_0)}
\def \sigG {\Sigma_{G, p_0}}

\usepackage{empheq}
\newcommand*\widefbox[1]{\fbox{\hspace{2em}#1\hspace{2em}}}
\newcommand{\magr}[1]{\textcolor{red}{#1}}

\reversemarginpar

\def\nn{\nonumber}

\def\l{\left}
\def\r{\right}
\def\DM{\mathrm{d}}
\def\lp{\ell_0}

\begin{document}
\title{
{Synge's World function and the quantum spacetime}
}
\vspace{0.25cm}

\author{Dawood Kothawala}

\address{
Centre for Strings, Gravitation and Cosmology, Department of Physics,
Indian Institute of Technology Madras, Chennai 600036, India
}

\ead{dawood@iitm.ac.in}

\begin{abstract}
All our observations that characterise space and time are expressed in terms of non-local, bi-tensorial objects such as geodesic intervals between events and two-point (Green) functions. In this contribution, I highlight the importance of characterising spacetime geometry in terms of such non-local objects, focusing particularly on two important bi-tensors that play a particular fundamental role -- Synge's World function and the van Vleck determinant. I will first discuss how these bi-tensors help capture information about spacetime geometry, and then describe their role in characterising quantum spacetime endowed with a lower bound, say $\lp$, on spacetime intervals. 
Incorporating such a length scale in a Lorentz covariant manner necessitates a description of spacetime geometry in terms of above bi-tensors, and naturally replaces the conventional description based on the metric tensor $g_{ab}(x)$ with a description in terms of a non-local bi-tensor $q_{ab}(x, y)$. The non-analytic structure of $q_{ab}(x, y)$ which renders a perturbative expansion in $\lp$ meaningless, also generically leaves a non-trivial ``relic" in the limit $\lp \to 0$. I present some results where such a relic term is manifest; specifically, I will discuss how this: (i) suggests a description of gravitational dynamics different from the one based on Einstein-Hilbert lagrangian, (ii) implies dimensional reduction to $2$ at small scales, (iii) connects with the notion of cosmological constant itself being a non-local vestige of the small scale structure of spacetime, (iv) helps address the issues of spacetime singularities.
I will conclude by discussing the ramifications of these ideas for quantum gravity.
\end{abstract}

\pagebreak
 
\hrule width \hsize \kern 1mm \hrule width \hsize height 2pt 

\vspace{0.2cm}

{\fontencoding{OT1}\fontfamily{ppl} 
\noindent Now it seems that the \textit{\textbf{empirical notions on which the metrical determinations of space are founded}}, the notion of a solid body and of a ray of light, \textit{\textbf{cease to be valid for the infinitely small}}. We are therefore quite at liberty to suppose that the metric relations of space in the infinitely small \textit{\textbf{do not conform to the hypotheses of geometry}}; and we ought in fact to suppose it, if we can thereby obtain a simpler explanation of phenomena. 

\vspace{0.2cm}

\noindent The question of the \textit{\textbf{validity of the hypotheses of geometry in the infinitely small}} is bound up with the question of the ground of the metric relations of space \ldots 

\vspace{0.2cm}

\noindent The answer to these questions can only be got by starting from the conception of phenomena which has hitherto been justified by experience, and which Newton assumed as a foundation, and \textit{\textbf{by making in this conception the successive changes required by facts which it cannot explain}} \ldots

\vspace{0.2cm}

\noindent \textit{\textbf{This leads us into the domain of another science, of physic, into which the object of this work does not allow us to go today.}} 

}

\vspace{0.25cm}

\hfill {\it from}: \textbf{On the Hypotheses which lie at the Bases of Geometry}, 

\hfill \textbf{Bernhard Riemann}, Gottingen lecture, 1854  (translated by W. Clifford)


\vspace{0.2cm}

\hrule width \hsize \kern 1mm \hrule width \hsize height 2pt 

\section{Introduction}
\phantom{blah}

Riemann, in his classic lecture on the foundations of geometry, clearly highlights the importance of understanding space in the context of physical effects and observations, while alerting that our usual conceptions of space might not hold at very small scales. The statements quoted above are particularly prescient, since they it is only in recent times that studies in quantum gravity have revealed how true Riemann's insights were. All our physical measurements are characterised in terms of extended solid objects and light rays ("rods and clocks") which we employ as probes, and hence, it is but to be expected that the quantum behaviour of these probes must play an important role in characterising the structure of spacetime. In brief:
\\
\\
\textit{We should be able to re-construct spacetime purely in terms of quantities that characterise such measurements.}
\\
\\
While easy to state, this basic, fundamental fact has been forgotten in the long history of gravitational physics, in which our geometric description of gravitational dynamics has been rooted in terms of local, tensorial quantities such as $g_{ab}(x), R_{abcd}(x), etc.$. While such quantities can, and do, characterise motion and measurements, they do not capture the essence of the fact highlighted above: spacetime must be re-constructed from observables tied to measurements. This shift of viewpoint, which might seem to be a matter of taste at the classical level, becomes absolutely essential at the quantum level where local, tensorial objects might not make much sense. In fact, if the quantum effects are non-analytic, then the limit in which one recovers locality might not commute with the limit $\hbar \to 0$, and hence: 
\\
\\
\textit{Even the classical limit might carry some vestige of an inherently quantum spacetime.}
\\
\\
In this contribution, we will describe a formalism attempts to re-construct spacetime in terms of non-local bi-tensors, $Q_{ab \ldots i'j' \ldots}(x,x')$, with the (un-)primed indices referring to an event with coordinate $x$ ($x'$) in some suitable chart. In particular, I will first describe how classical metric can be described in terms of such quantities, and then try to obtain an "effective metric" for the quantum spacetime by imposing the condition that there exists a lower bound on spacetime intervals. The brief plan of this contribution is:
\begin{itemize}
\item {\bf Sec 2:} Small scale structure of spacetime -- the zero point length
\item {\bf Sec 3:} Reconstructing spacetime geometry from Synge World function -- replacing the notion of ``metric" with two-point functions
\item {\bf Sec 4:} Quantum aspects of the reconstructed spacetime
	\begin{itemize}
	\item Relics of the quantum spacetime -- the limit $\lp \to 0$ 
	\item Effective dimensional reduction at small scales
	\item Spacetime singularities
        \item Causal structure and ``silence" at small scales
	\end{itemize}
\item {\bf Sec 5:} Emergent gravity as a relic of quantum spacetime
\item {\bf Sec 6:} A broader outlook -- bi-tensors as tools to probe classical and quantum spacetime
\end{itemize}
\section{Small scale structure of spacetime 
\\
\textcolor{white}{bl} -- the \textit{zero point} length of spacetime}
$$$$
In almost any attempt to combine the principles of quantum mechanics and general relativity, there are atleast two results, rooted in semi-classical formalism but widely considered robust to the new physics which the domain of quantum gravity might bring in. These are: 
\\
\\
(a) Thermal nature of acceleration horizons (Unruh effect and Hawking radiation), and 
\\
(b) Existence of a lower bound to spacetime intervals -- a minimal length scale -- which arises very naturally when one tries to combine the principles of quantum mechanics and general relativity. 
\\
\\
We will largely be interested in (b) in this contribution, but in doing so, will be led to an intriguing connection with one of the implications of (a) - the emergent paradigm of gravity. We will discuss this connection towards the end. The most basic manner in which (b) can be introduced in an effective manner into the structure of spacetime - without resorting to any specific framework of quantum gravity - is to change focus from a description in terms of metric tensor, and instead work directly with the distance function $d(x,y)$ between spacetime events. Existence of a minimal length scale can then be introduced in terms of the modifications of the distance function $d(x,y)$ between spacetime events. However, in Lorentzian geometry, a better variable to use is the squared geodesic interval $\sigma(x,y)^2$, which is one half of the so called Synge world function $\Omega(x,y)$. This is related to the distance function by $d(x,y)=\sqrt{|\sigma(x,y)^2|}$. The existence of a lower bound to spacetime intervals can then be stated in a very simple manner: The presence of a lower bound to geodesic intervals manifests itself as the modification of the squared geodesic interval
	\begin{eqnarray}
	\sigma^2 &\to& \mathcal{S}(\sigma^2)
	\nn \\
	\mathcal{S}(0) &=& \pm \lp^2 \neq 0
	\end{eqnarray}
This is the most basic, Lorentz invariant, statement that in the coincidence limit $y \to x$, $\sigma^2 \to 0$, the modified geodesic distance should be non-zero. 
\footnote{As an aside, let us point out that any such modification would generically violate the axioms of metric spaces. This is not unexpected and, in fact, it is not difficult to show that even the equality part of the triangle inequality can not hold in presence of a minimal length. The key reason for this is clear - the condition $\mathcal{S}(0) \neq 0$ essentially implies that one can not localise an event to an accuracy better than $\lp$. Therefore, given an ordered sequence of points $p$, $q$, and $r$ on a geodesic, a statement about geodesic distances between these can not hold to an accuracy better than $\lp$, since none of these points can be localised to that accuracy.
}
%

Given the above simple fact, how far can one go towards constructing variables analogous to the metric tensor $g_{ab}(x)$ which we conventionally use to describe spacetime? The answer to the above question lies in the perhaps not so well appreciated importance of the world function, $\Omega(x,y)=\frac{1}{2} \sigma(x,y)^2$: much of the information about $g_{ab}(x)$, $R_{abcd}(x)$, etc.  can be derived from coincidence limits of derivatives of $\Omega(x,y)$. This has important consequence, which will become apparent as we proceed.
\\
\\
\noindent \underline{\textit{Notation}:} 
We work in $D$ dimensions, and use the sign convention $(-, +, +, \ldots)$ for Lorentzian spaces. Latin alphabets denote spacetime indices. Also, for notational convenience, we use $\lp^2$ throughout to denote short distance cutoff on geodesic distances; for timelike/spacelike cases, the replacement $\lp^2 \rightarrow \epsilon \lp^2$ must be made in the final results after which $\lp^2>0$. For convenience, we give below a quick list of some of the most recurring symbols/notation used in the text: We use $D_k \overset{\rm def}{: \Longrightarrow} D-k$, where $D$ is the dimension of spacetime, as a convenient shorthand. $p_0$ represents a fixed spacetime event (the {\it base point}), and $p$ the variable event (the {\it field point}). Occasionally, coordinate labels $y, x$ are used to represent $p_0, p$ respectively. $\Rsg$ denotes the Ricci scalar; similarly $\Rsq$ denotes the Ricci bi-scalar built from the qmetric (see text). $\mathcal S_{ab} = R_{ambn} q^m q^n$; $\mathcal S = g^{ab} \mathcal E_{ab} = R_{ab} q^a q^b$. Finally, $\l[\Rsqn\r](p_0)$ represents the coincidence limit of $\Rsq$.
\\
\\
\noindent \underline{\textit{References}}: Before proceeding to describe the framework, results, and implications based on the above idea, I give below a brief survey of the references {\it vis a vis} the various results they contain. \textit{\textbf{No explicit citations will be given in the main body of the text unless absolutely required. Also, the {\it Bibliography} at the end contains references describing the formalism reviewed here.}} The reader is referred to these papers for further references to important papers directly and/or indirectly relevant to this work.

\begin{itemize}
\item A framework based on the above was first described in Ref.~\cite{dk-ml}, where the key mathematical inputs and conceptual points were discussed in detail. 

\item A generalisation of this was subsequently studied in Ref.~\cite{dk-tp-ml}, which presented the important result for the Ricci bi-scalar $\Rsq$ constructed from $q_{ab}(x;y)$. In particular, it was pointed out that the limit $p \to p_0$ of $\Rsq$ is non-trivial in it's dependence on $\lp$. A similar computation for the Gibbons-Hawking-York (GHY) boundary term was presented in Ref.~\cite{dk-tp-ghy}. Ref.~\cite{dk-tp-ml} in particular presented and highlighted the conceptual importance of the results concerning minimal length and qmetric for the {\it Emergent Gravity} paradigm and the {\it Cosmological Constant} problem. It also provides a broader perspective on the entire framework, with examples from several other areas of theoretical physics that are helpful in better understanding of the result.

\item Perhaps the most important missing term in the above computations was the dependence of $q_{ab}(x;y)$ on the van Vleck determinant $\Delta(x,y)$. This was studied in detail in Ref.~\cite{jaffino-dk}, which presents all the derivations and results in a mathematical rigorous manner, and derives $q_{ab}(x;y)$ in it's final form. 


\item This final form was then employed in Ref.~\cite{sc-tp-dk} to study the volumes and areas of equi-geodesic surfaces of ``radius" $\sigma(x;0)^2=R$ using $q_{ab}(x;0)$, leading to yet another important result: The effective dimension $D_{\rm eff} \to 2$ in the limit $x \to 0$, regardless of the dimensionality $D$ of the original spacetime.

\item Implications of the qmetric for caustics of geodesic congruences, via Raychaudhuri equation, was discussed in Ref.~\cite{sc-dk-pesci}.
\end{itemize}

\noindent I will now describe and summarise these results, their implications, and future outlook.

\section{Reconstructing spacetime geometry from Synge World function
\\
\textcolor{white}{bl}  -- replacing the notion of ``metric" with two-point functions}
$$$$

\subsection{Geodesic distances $\sigma(x,y)^2$ as more fundamental than the metric $g_{ab}(x)$}

\begin{figure}[H]%
	    \centering
	    \subfloat[Conventional expansion of the metric near an event $p_0$.]{{\includegraphics[width=0.525\textwidth]{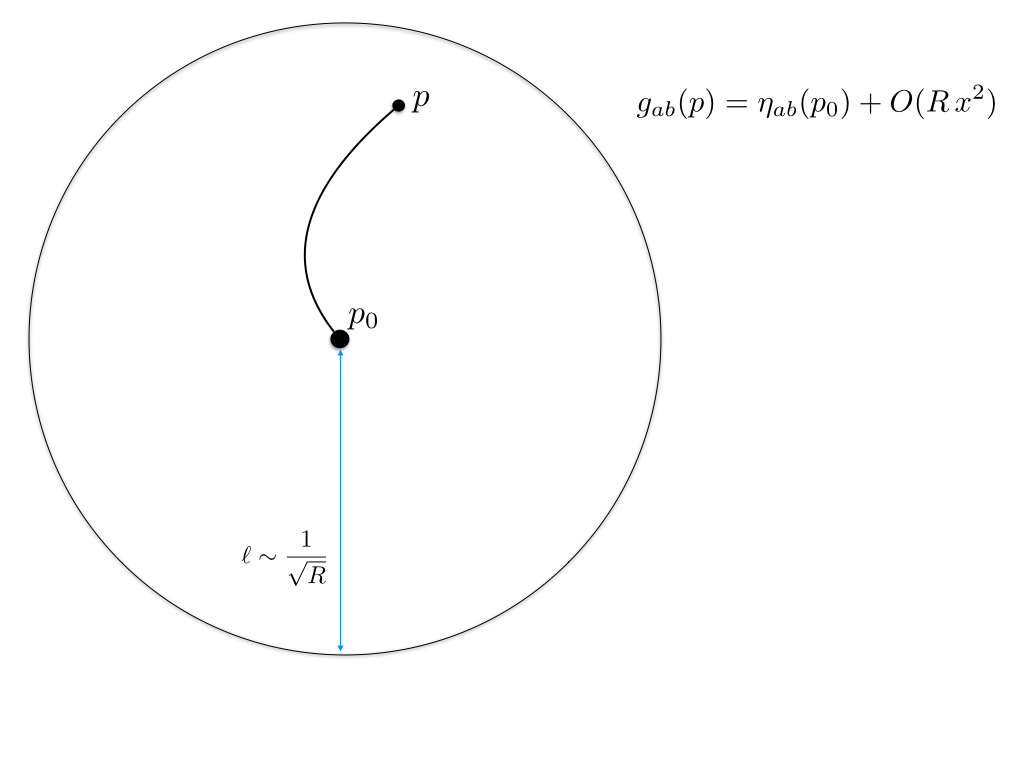} }}%
	    \subfloat[The qmetric at small scales.]{{\includegraphics[width=0.525\textwidth]{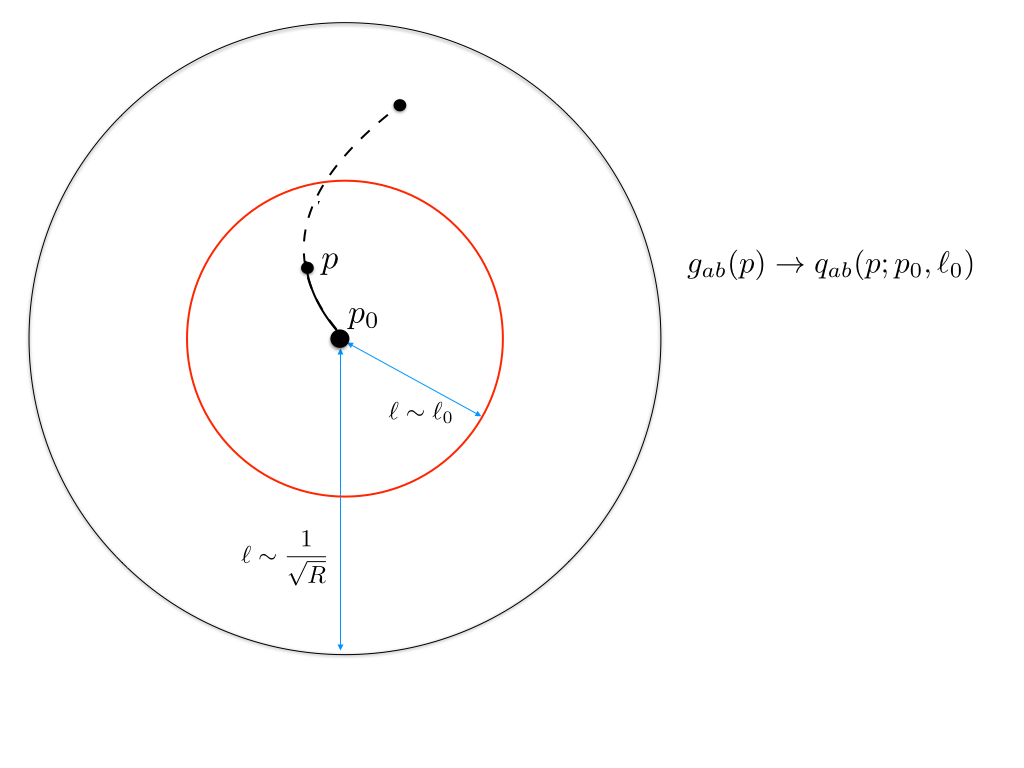} }}%
	    \caption{The modification of conventional metric description by the bi-tensor $q_{ab}$ at small scales.}%
	    \label{fig:scales}%
	\end{figure}

As mentioned above, our key idea is to replace the description of spacetime in terms of the metric tensor, with one that uses geodesic distances as the key variables. The possibility of doing so is based on the existence of mathematical identities relating the local tensorial quantities (metric curvature, etc.) that characterise a spacetime to the coincidence limit (denoted by ``$[\ldots]$") of derivatives of $\sigma(x,y)^2$. For example, in terms of Synge's world function $\Omega(x,y)=(1/2) \sigma(x,y)^2$, we have 
	\begin{eqnarray}
	g_{a'b'} = g_{ab} &=& \l[{\nabla}_a {\nabla}_b \Omega(x,x') \r] = \l[{\nabla}_{a'} {\nabla}_{b'} \Omega(x,x') \r] 
	\nn \\
	R_{a' (c' d') b'} &=& \l(3/2\r) \l[{\nabla}_a {\nabla}_b \nabla_{c} \nabla_{d} \Omega(x,x') \r]
	\end{eqnarray}
As we argued above, since a Lorentz invariant lower bound is best introduced through modification of geodesic distances, it then makes sense to ``reconstruct" the ``metric" from the distance function. While this procedure gives the conventional metric tensor $g_{ab}(x)$ in absence of $\lp$, we shall see that it leads to a non-trivial bi-tensor $q_{ab}(p;p_0)$ when $\lp$ is introduced. It is this object which we shall call as the {\it qmetric}, and use as the key variable in our description of small scale structure of spacetime. Before proceeding to discuss the qmetric, it is worth emphasizing at this stage why the conventional Taylor expansion of the metric in Riemann normal coordinates (RNC):
\begin{eqnarray}
g^{\tiny{\rm RNC}}_{ab}(x;y) = \underset{\magr{\bm ?}}{{\bm \eta_{ab}(y)}}  -  \frac{1}{3} \underset{\magr{\bm ?}}{{\bm R_{a c b d}(y)}}~(x-y)^c (x-y)^d + \mathrm{higher~order~terms}
\label{eq:rnc}
\end{eqnarray}
is not expected to hold for points $x$ that lie within a geodesic distance of the order of $\lp$ from $y$. First, it is not clear whether the leading term would be Minkowski metric tensor (due to quantum fluctuations). Moreover, such an expansion would break down near a point $y$ if any of the basis components of the Riemann tensor diverge at $y$. Since a minimal length scale is expected to have a quantum gravitational origin, and since our qmetric is derived assuming such a length scale exists, we may therefore not expect that 
\begin{eqnarray}
q_{ab}(x;y) \not\equiv g^{\tiny{\rm RNC}}_{ab}(x;y)
\end{eqnarray}
This is illustrated in Fig. \ref{fig:scales}. Indeed, as it will turn out, $q_{ab}(x;y)$ has a much more non-trivial structure.

\subsection{Geodesic structure of spacetime}
Since we are proposing to use geodesic distance as the key variable, it is natural to expect the mathematical construction of the qmetric would be based on the geodesic structure of spacetime. This is characterized by the what we shall call as the  {\it equi-geodesic} surfaces $\sigG \equiv \{ p \in \mathcal N| \Omega(p,p_0)= \rm constant \}$ in 
a geodesically convex neighbourhood $\mathcal N$ of an event $p_0$. We begin by discussing the intrinsic and extrinsic geometry of {\it equi-geodesic} surfaces $\sigG$. This is illustrated in Fig. \ref{fig:geodesic-congruence}.
{
	\begin{figure}[!htb]%
	    \centering
	    \subfloat[Equi-geodesic surfaces $\sigma_G$ attached to an event $p_0$ in an arbitrary curved spacetime.]{{\includegraphics[width=0.6\textwidth]{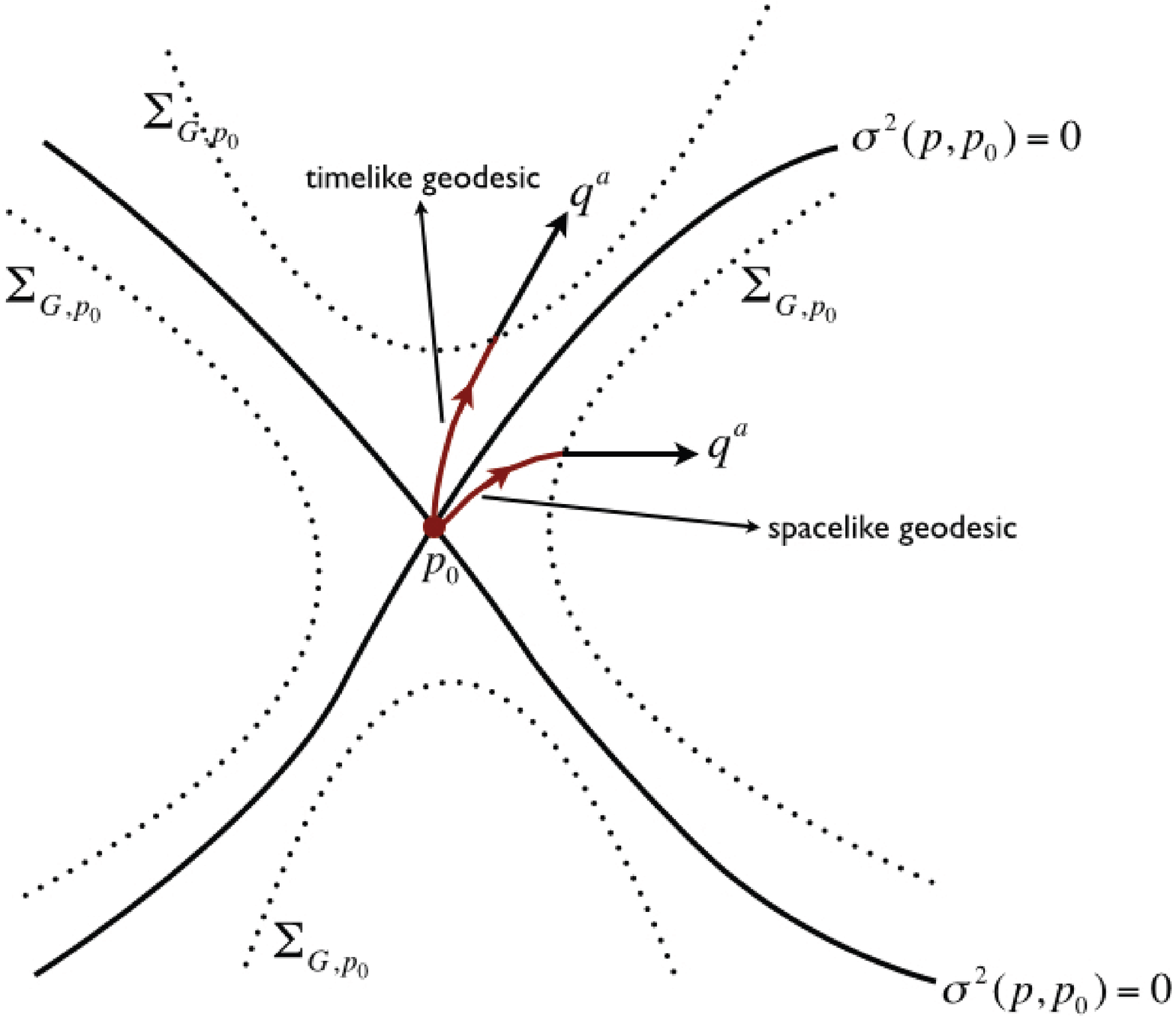} }}%
	    \subfloat[$\sigma_G$ in Minkowski spacetime.]{{\includegraphics[width=0.4\textwidth]{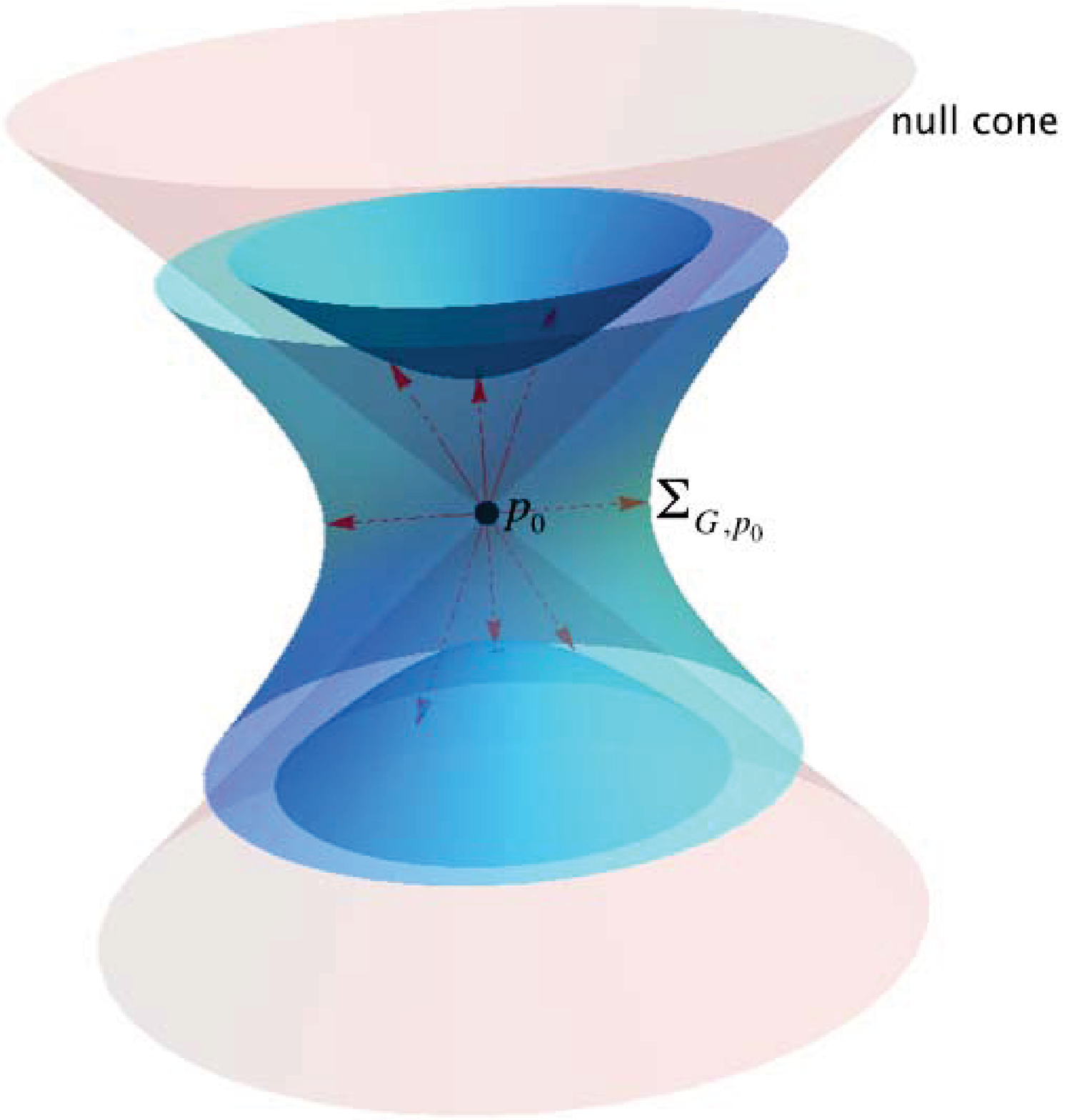} }}%
	    \caption{The geodesic structure of spacetime.}%
	    \label{fig:geodesic-congruence}%
	\end{figure}
}
The intrinsic and extrinsic geometry of these surfaces are characterised by their intrinsic curvature scalar $R_{\Sigma}$ and the extrinsic curvature $K_{ab}$. These are the quantities that will appear in the Ricci bi-scalar $\Rsq$ constructed from the qmetric. Moreover, in smooth regions of the background spacetime (that is, away from curvature singularities), the following expansions will turn out to be useful:
\begin{eqnarray*}
K_{ab} &=& \frac{1}{\lambda}h_{ab} - \frac {1} {3} \lambda \mathcal{S}_{ab} + \frac {1} {12} \lambda^2 \nabla_{\bm t} \mathcal{S}_{ab} - \frac {1} {60}\lambda^3 F_{ab} + O(\lambda^4)
\nn \\
\nn \\
K &=& \frac{D_1}{\lambda} - \frac {1} {3} \lambda \mathcal{S} + \frac {1} {12} \lambda^2 \nabla_{\bm t} \mathcal{S} - \frac {1} {60}\lambda^3 F + O(\lambda^4)
\nn \\
\nn \\
R_{\Sigma} 
&=& \frac {\epsilon D_1 D_2} {\lambda^2} + R - \frac{2\epsilon(D+1)}{3} \mathcal{S} + O(\lambda)
\end{eqnarray*}
where: $\mathcal{S}_{ab}=R_{aibj} t^i t^j$, $\mathcal{S}=R_{ab} t^a t^b$ and $F_{ab} = \nabla_{\bm t}^2 \mathcal{S}_{ab} + (4/3) \mathcal{S}_{ak} \mathcal{S}^k_{\phantom{k}b}$, with $F = F_{ab} g^{ab}$.

\subsection{{\it qmetric}: the final form}
We now put together our key inputs, collect certain mathematical identities related to the above mentioned bi-tensors, and attempt to construct the qmetric. Let us first state the inputs clearly, before implementing them mathematically:

\vspace{0.5cm}

\noindent {\bf Q1:} Geodesic distances are modified so that they have a Lorentz invariant lower bound. As already stated above, this is summarised by 
$\sigma^2 \rightarrow  {\mathcal S}\l[ \sigma^2 \r]$ with ${\mathcal S}\l[ 0 \r] = \lp^2$.

\vspace{0.25cm}

\noindent The precise details of the function $\mathcal S\l[ \sigma^2 \r]$ must come from a complete framework of quantum gravity, and we do not impose or choose any particular form, except for demonstrating certain qualitative features of the results. 
\footnote{We must, however, state that there exist several motivations from older works that selects the form $\mathcal S(z) = z + \lp^2$.} 

\vspace{0.5cm}

\noindent {\bf Q2:} The modified d'Alembartian $\widetilde{ _{p_0} \square_p}$ yields the following modification for the two point functions $G(p,p_0)$ of fields in {\it all maximally symmetric spacetimes}: $G\l[\sigma^2\r] \rightarrow \widetilde G\l[\sigma^2\r] = G\l[ {\mathcal S}_{\lp}\l[ \sigma^2 \r] \r]$

\vspace{0.25cm}

\noindent We have not discussed this particular input before, but it comes mainly from the fact that existence of a minimal length is expected to regularise the divergences in QFT, and, in particular, modify the structure of the two-point functions. We refer the reader to the references for further discussion on this point. Here, we only point out one immediate implication of the same. Since the leading form of the two point function in any spacetime is given by  	
	\begin{eqnarray}
	G(x,y) := \frac{\sqrt{\Delta(x,y)}}{\l(\sigma(x,y)^2\r)^{\frac{D-2}{2}}} \times \l( 1 + \mathrm{subdominant~terms} \r)
	\nn
	\end{eqnarray}
this second input clearly selects the bi-tensors $\Delta(x,y)$ and $\sigma(x,y)^2$ on which we can expect the qmetric to depend.

\vspace{0.5cm}

\noindent I will now sketch how the above inputs can be implemented.
\begin{enumerate} 
\item To implement {\bf Q1}, we use the defining Hamilton-Jacobi eq for the world fn 
\begin{eqnarray}
g^{ab} \; \partial_a \Omega \; \partial_b \Omega = 2 \Omega
\,\,\, \longrightarrow \,\,\,
q^{ab} \partial_a \mathcal S_{\lp} \partial_b \mathcal S_{\lp} = 4 \mathcal S_{\lp}
\nn \label{eq:hjq}
\end{eqnarray}

\item To implement {\bf Q2}, we use the following non-local expression for the d' Alembartian $\widetilde \square$ constructed from $q_{ab}(x;y)$
{\normalsize
\begin{eqnarray}
\label{boxq}
\widetilde \square = A^{-1} \Biggl\{ \square_g+\frac{1}{2}D_3~g^{ij}\partial_i \ln A~\partial_j + \epsilon \slashed{\partial}\ln A~\slashed{\partial} \Biggl\}
&+& \epsilon Q \Biggl\{ \l[ \nabla_i q^i + \frac{1}{2}D_1\slashed{\partial}\ln A\r]\slashed{\partial}+\slashed{\partial}^2 \Biggr\}
+
\sqrt{\epsilon \sigma^2}\alpha'\slashed{\partial}
\nn
\end{eqnarray}
}
(where $\slashed{\partial}=q^i \partial_i$) and following identities satisfied by the VVD
\begin{eqnarray}
\Delta(p,p_0) = \frac{1}{\sqrt{|g(p)|} \sqrt{|g(p_0)|}} {\rm det} \l\{ \overset{(p)}{\nabla_a} \overset{(p_0)}{\nabla_b} \; \Omega(p,p_0) \r\}
\nn
\end{eqnarray}

\begin{eqnarray}
I1:&& \hspace{0.1cm} \nabla_{\bm q} \ln \Delta = \frac{D_1}{\sqrt{\epsilon \sigma^2}} - K
\nn \\
I2:&& \hspace{0.1cm} \nabla_{\bm q}^2 \ln \Delta = - \frac{D_1}{\epsilon \sigma^2} + K_{ab}^2 + R_{ab} q^a q^b
\label{eq:VVD-ids}
\nn
\end{eqnarray}
\end{enumerate}

Using the above, the qmetric can now be derived, though the details of the derivation are lengthy and can be found in the references, specifically Ref.~\cite{jaffino-dk}.
\begin{eqnarray}
q_{ab}(x;y) &= \frac{\mathcal S_{\lp}}{\sigma^2} \l(\frac{\Delta_{\phantom{\mathcal S}}}{\Delta_{\mathcal S}}\r)^{+\frac{2}{D_1}} g_{ab}(x) \; + \; 
 \epsilon \Biggl\{ \frac{\sigma^2 \mathcal S_{\lp}'^2}{\mathcal S_{\lp}} - \frac{\mathcal S_{\lp}}{\sigma^2} \l(\frac{\Delta_{\phantom{\mathcal S}}}{\Delta_{\mathcal S}}\r)^{+\frac{2}{D_1}} \Biggl\} \; t_a(x;y) \; t_b(x;y)
\nn
 \label{eq:qmfinal}
\end{eqnarray}
where $t_a = g_{ab} q^b$ (see Fig. \ref{fig:geodesic-congruence}). Note that the qmetric is:
	\begin{itemize} 
	\item[\tiny $\bullet$] A {non-local bi-tensor}, {disformally coupled} to $g_{ab}$
	\item[\tiny $\bullet$] {Singular} in the limit $\sigma^2 \to 0$
	\item[\tiny $\bullet$] $\lim \limits_{\lp \to 0} q_{ab}(x;y) = g_{ab}(x)$
	\end{itemize}

\noindent As an important point, we must highlight that although the qmetric above has been constructed in a timelike or spacelike neighbourhood of any event $p_0$, there does exist a generalisation for events separated from $p_0$ along null intervals, by A. Pesci; this can be found in Refs. \cite{qmetric-null-pesci}. Before we discuss some implications of this work, let me address a couple of often-asked questions:
\\
\\
\noindent {\it Question 1}: {How generic will such a result be as far as quantum gravity is concerned}?
\\
\noindent The key insight coming from our analysis, going beyond the specifics, is the following: It seems likely that non-local but (hopefully) Lorentz in(co)variant deformation of the spacetime geometry at small scales, will be a consequence of quantum fluctuations of the (as yet unknown) microscopic degrees of freedom of quantum gravity.  This leads to an inevitable $O(1)$ modification due to a minimal spacetime length. Regardless of the precise form of the deformation, the results will now essentially depend on the two limits, $\xi \to 0$ and $\xi \to \infty$ (with $\xi=\sigma^2/\lp^2$) being inequivalent. Any dimensionless deformation function $A[\sigma, \lp]$ can only depend on $\xi$; $A[\sigma, \lp]=A[\xi]$. Hence, unless $A[\xi]$ has the same limit at $\xi \to 0$ and $\xi \to \infty$, (e.g., it is symmetric under $\xi \to 1/\xi$) one will generically obtain the kind of non-trivial result that we have obtained.
\\
\\
\noindent {\it Question 2}: {To what extent can we think of the qmetric as a metric}?
\\
\noindent This issue is somewhat irrelevant to our results and -- more generally -- when one treats a distance function $d(x,y)$ defined on a manifold as more fundamental than the metric, and the metric itself as a derived quantity which enables us to construct geometric invariants for a manifold. As explained in detail in Ref.~\cite{dk-ml}, the non-local character of the qmetric is simply a physical characterization of quantum fluctuations which leave their imprint (in this approach) in the form of a lower bound on the intervals:
$d(x,y) \geq \lp$. At the smallest of the scales, we do not expect any local tensorial object to describe the quantum spacetime geometry accurately anyway, and hence it is not unexpected that more general mathematical objects must replace the conventional ones. Our work provides a first, mathematically concrete, step in this direction. 
In fact, non-local effective actions have been considered for quite some time in the context of quantum gravity (e.g., DeWitt proposed such an action in 1981 \cite{dewitt}), and $q_{ab}(p, p_0)$ might serve as an important mathematical object to build such actions. That is all we need as far as physics is concerned.

\section{Quantum aspects of the reconstructed spacetime}
\phantom{blah}

Having described the geometrical construction, we now highlight the key implications and results that follow naturally from the reconstructed metric.

\subsection{Relics of the quantum spacetime -- the limit $\lp \to 0$}
$$$$
Given the singular behaviour of $\bm q$, it is unclear whether {local scalars} constructed out of $q_{ab}$ reduce, in the limit $\lp \to 0$, to their corresponding form in $\bm g$. Therefore, the question of interest is whether the limit $\lp \to 0$ is a singular limit of a fundamentally non-local formalism. One can expect to get interesting and non-trivial, insights into the classical theory if 
$$
\lim \limits_{\lp \to 0} \lim \limits_{y \to x} \neq \lim \limits_{y \to x} \lim \limits_{\lp \to 0}
$$
Going back to the question of $\lp \to 0$ limit of ``{local scalars}" constructed out of $q_{ab}$, let us start by constructing the simplest geometrical quantity - the Ricci ``scalar":
	\begin{eqnarray}
	 \l[ \Rsqn \r](p_0) \overset{?}{=} \Rsg + {\rm terms~of~order~} \lp
	\nn
	\end{eqnarray}
Is the answer to the ``?" above yes? If so, one would have proved that the coincidence limit of the Ricci bi-scalar constructed from the qmetric has an expansion in $\lp$ whos leading term is the Ricci scalar of $g_{ab}(x)$. If the limits highlighted in the title of this section were the same, this would indeed be the case. This, however, turns out not to be the case! In fact, not only do we have $${\lim \limits_{\lp \to 0} \lim \limits_{\sigma^2 \to 0}} \Rsq \neq \Rsg$$ but what we do get for these limits turns out to have important implications for the emergent gravity paradigm.

The {\bf exact} form of the Ricci scalar can be written in a compact form in terms of geometric quantities associated with 
{$\sigma^2=$ const surface, $\Sigma$}
\vspace{.2cm}

{{
\begin{eqnarray}
\label{eq:finalRsq}
      \Rsq =
      \Biggl[ \frac{\sigma^2}{\mathcal S_{\lp}} \zeta^{-{2}/{D_1}} \; \mathcal R_{\sigG} - \frac{D_1 D_2}{\mathcal S_{\lp}} + 4 (D+1) (\ln \Delta_{\mathcal S})^{\bullet} \Biggl] 
-			\frac{\mathcal S_{\lp}}{\magr{\lambda^2} \mathcal S'^2_{\lp}}  \Biggl\{ K_{ab} K^{ab} - \frac{1}{D_1} K^2 \Biggl\}
			\nn \\
			\;+\; {4 \mathcal S_{\lp}  \Biggl\{ - \frac{D}{D_1} \l[ (\ln \Delta_{\mathcal S})^{\bullet} \r]^2 + 2 (\ln \Delta_{\mathcal S})^{\bullet\bullet} \Biggl\}}
			\nn
\end{eqnarray}
}}
\\
where $\Delta_{\mathcal S}$ is defined as $\Delta$ with $\sigma^2$ replaced by $\mathcal S$, and
$\zeta={\Delta}/{\Delta_{\mathcal S}}$, 
$(\ln \Delta_{\mathcal S})^{\bullet} = \DM \ln \Delta_{\mathcal S}/\DM \mathcal S_{\lp}$, 
$(\ln \Delta_{\mathcal S})^{\bullet\bullet} = \DM(\ln \Delta_{\mathcal S})^{\bullet}/\DM \mathcal S_{\lp}$.

We emphasize that the above expression is exact, with no Taylor expansions anywhere. However, we now analyse its behaviour in a smooth region of spacetime, where the Taylor expansions of various quantities associated with the equi-geodesic surfaces given above will hold. After lengthy algebra, one obtains

\begin{subequations}
\begin{empheq}[box=\widefbox]{align}
	\l[\Rsqn\r](p_0) = \underset{p \rightarrow p_0}{\lim} \Rsq
	= \underbrace{\alpha \l[R_{ab} t^a t^b \r]_{p_0}}_{O(1) \; \rm term} 
\; + \;
{\lp^2} \;
\underbrace{
\l[~\rm curvature~squared~terms~ \r]_{p_0}
}_{O(\lp^{\;2}) \; \rm term}
 \nn
\end{empheq}
\end{subequations}
thereby giving a non-trivial limit.
{
{{Although $q_{ab} \to g_{ab}$ when $\lp=0$, $\l[\Rsqn\r](p_0) \neq \Rsg$ in the same limit! 
\\
\\
\textit{The ``zero-point length" leaves it's vestige \ldots much like the grin of the Cheshire cat!}
\\
\\
{Most importantly, the leading term above is precisely the {entropy density} which arises as {Noether charge} of diff-invariance, and plays a prominent role in the {\it Emergent gravity} paradigm!}
}}
} An exactly similar analysis can be done for the GHY surface term in the Einstein-Hilbert action, and can be found in Ref.~\cite{dk-tp-ghy}, and we shall not discuss it here. 

\vspace{0.25cm}

\noindent Instead, we briefly comment on the conceptual implications of the fact that the limit $\underset{\lp \rightarrow 0}{\lim} \underset{p \rightarrow p_0}{\lim}$ for certain observable $\mathcal{O}_q$, constructed using the qmetric, may not correspond to the classical expression for $\mathcal{O}_g$. Our result for the Ricci scalar demonstrates the possibility of this happening, and we would like to briefly highlight and put in context an important conceptual issue that arises from this:  Our result suggests that {non-local} and {non-analytic} effects of a minimal length might leave residues which are independent of $\lp$: {\it How far can we trust such relics}? In fact, {quantum relics} of similar nature are not unfamiliar in physics; for e.g. effects of Lorentz violating regulators at higher energies can generically get dragged to lower energies due to radiative corrections, leaving $O(1)$ residual effects {[Collins et. al., Polchinski] \cite{collins-etal}}; conformal anomaly; $D\to4$ limit in dimensional regularization. {[see, for e.g., Birrell \& Davies]}; non-relativistic relic from $c^{-1}$ expansion of the relativistic point particle wave fn {[Padmanabhan et. al.] \cite{paddy-hamsa}}; non-local quantum residue of discreteness of the causal set type {[Sorkin] \cite{sorkin}}. These examples show that when certain limits are taken in a theoretical model the resulting theory could contain relics of the more exact description. In all such cases no amount of study of the approximate theory will give us a clue as to where the relic came from (e.g., the study of Schrodinger equation for the helium atom can never lead us to the Fermi statistics for the electrons). We believe our result is of similar nature, and is nearly impossible to understand within the context of classical gravity itself. Our analysis throws light on this and shows that it could be a valid relic of quantum gravity. 
	
An immediate implication of this result would then be to see whether the relic term explains some phenomenon which is difficult to explain in the conventional framework. We will discuss this in the rest of this section.
%

\subsection{Effective dimensional reduction at small scales}
$$$$
The qmetric we have obtained captures the essence of existence of a minimal length scale, and it is therefore worth using it to explore geometrical quantities which could yield further information about the small scale structure of spacetime. 
	\begin{figure}[!htb]
	\scalebox{0.3}{\includegraphics{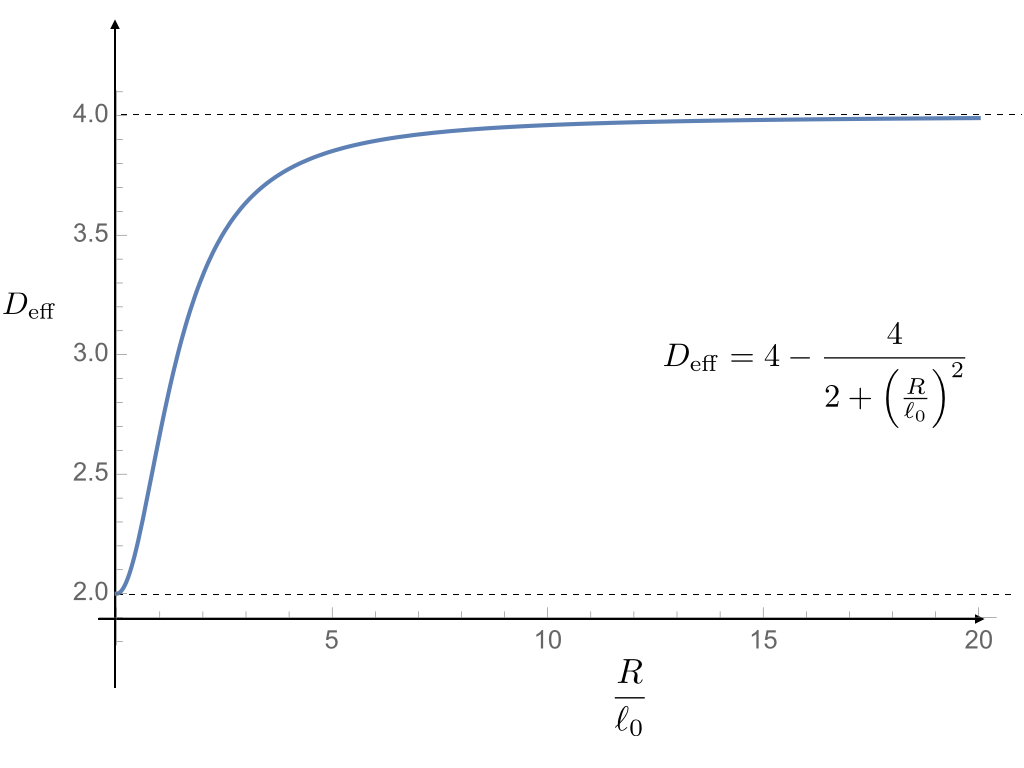}}
	\caption{The effective dimension associated with the qmetric goes from $D$ at large scales to $2$ at small scales.}%
	\label{fig:Deff}%
	\end{figure}
The key idea is to compute the area of, and spacetime volume bounded by, the equi-geodesic surfaces. This exercise was done in Ref.~\cite{sc-tp-dk}, where the following expressions were derived (for definiteness, the results were presented for the choice $\mathcal{S}(z)=z+\lp^2$) 
	\begin{eqnarray}
	\sqrt{q} &=& \sigma \l( \sigma^2 + \lp^2 \r)^{(D-2)/2} \l[ 1 - \frac{1}{6} \mathcal S_{\bm g} \l(\sigma^2 + \lp^2\r) \r] \sqrt{h_{\Omega}}
	\nn \\
	\nn \\
	\sqrt{h} &=& \l( \sigma^2 + \lp^2 \r)^{(D-1)/2} \l[ 1 - \frac{1}{6} \mathcal S_{\bm g} \l(\sigma^2 + \lp^2\r) \r] \sqrt{h_{\Omega}}
	\nn
	\end{eqnarray}
	where $ \mathcal S_{\bm g}=R_{ab} q^a q^b$ and $h_{\Omega}$ depends on coordinates on the equi-geodesic surfaces. We can now compute the volume 
	$V_D(R, \lp)$ of an qui-geodesic surface of size $R$: $\sigma^2 = R^2$ (in Euclidean), and read off the dimensionality of spacetime from the scaling of volume w.r.t. 
	$R$. For example, in the standard case $V_D(R,\lp=0) \propto R^D$. With this is mind, we can define an ``effective" {\it geometric} dimension as
	\begin{eqnarray}
	D_{\rm eff} = D + \frac{\DM}{\DM \ln R} \Biggl\{ \ln \l(\frac{V_D(R, \lp)}{V_D(R,\lp=0)}\r) \Biggl\}
	\nn
	\end{eqnarray}
	(The scaling with $V_D(R,\lp=0)$ is to mod out the effect of curvature, and does {\it not} affect the final result. See Ref.~\cite{sc-tp-dk} for details.) The above result blends in very well with the fact that several approaches to quantum gravity suggest such a dimensional reduction to $2$ at small scales. See Carlip \cite{carlip} for a review.

\subsection{Spacetime singularities}
$$$$
It is generally believed that the quantum theory of gravity must have something useful to say about the issue of curvature singularities that occur generically in classical general relativity. Since a consistent quantum theory of gravity is nowhere yet in sight, one can not attack the problem of singularity removal head on, but can take a cue from hints from general results in semi-classical gravity such as existence of a zero-point length. With this motivation, and as a first step, implications of the qmetric for the Raychaudhuri equation was analysed in \cite{sc-dk-pesci}, with two key results being: (i) existence of an upper bound on expansions of null and time-like geodesics, and (ii) additional terms in the Raychaudhuri equation related to the Van Vleck determinant associated with the modified geodesic interval.

A more direct approach to spacetime singularities is to construct the qmetric explicitly for a spacetime with curvature singularities, and see if the corresponding quantities are finite as one approaches the singularity. This program is complicated by two major factors: (i) Since curvature components would generically blow up near a singularity, one can no longer use Taylor expansions to analyse their local behaviour. One must instead use exact expressions for the World function and van Vleck determinant, which are not available. Nevertheless, some generic arguments can still be made, and specific results can be obtained for FLRW spacetimes. These indicate that (at least some) curvature quantities that blow up in the background spacetime are indeed rendered finite when constructed using the qmetric. These results are highly non-trivial in the sense that they involve considerable subtleties while taking the limits in presence of singularities, and can not be obtained by naive dimensional considerations (we already know $\lp$ disappears in the coincidence limit in a regular (non-singular) region of spacetime). A more powerful set of tools can be employed to address these and related issues \cite{dk-singularities}.

\subsection{Causal structure and ``silence" at small scales}
$$$$
One very straightforward and generic consequence of the non-local structure of the qmetric is to shrink the light cones at every event $p$ of spacetime, with respect to the light cones of the original metric. One may now define a suitable, covariant averaging scheme that takes into account such a shrinking of light cones due to all points $p_0 \in I^-(p)$ - the interior of the past light cone of $p$ - over a time scale $\tau$. One can give some very general arguments leading to specific results concerning the outcome of such an averaging for the causal structure associated with the qmetric. Several of these can be explicitly demonstrated for Minkowski and de Sitter spacetimes. One interesting fall-out of such an analysis is the possibility that the quantum spacetime, as described by the qmetric, might have a Euclidean regime at very small scales \cite{dk-singularities}.

	\begin{figure}[H]
	\begin{center}
	\scalebox{0.4}{\includegraphics{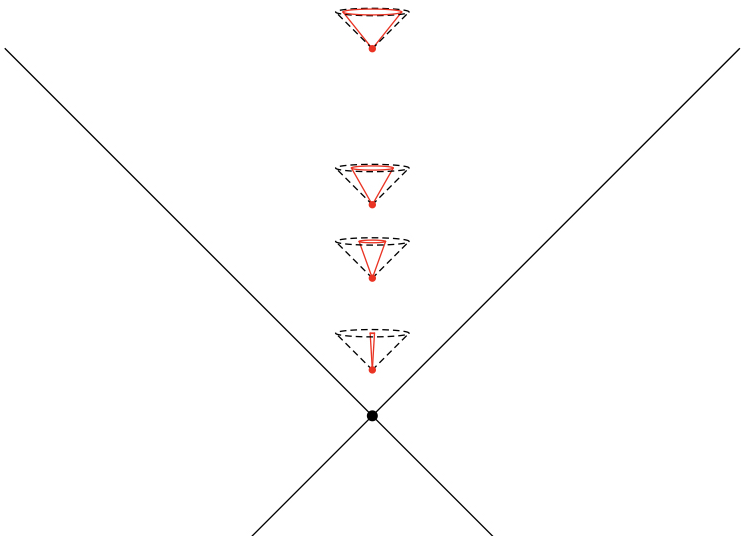}}
	\end{center}
	    \caption{Shrinking of null cones associated with qmetric (red) as compared to the light cones associated with the original metric (black). The base-point $p_0$ here 
	    is placed at the origin.}%
	    \label{fig:light-cones-1}%
	\end{figure}

\section{Emergent gravity as a relic of quantum spacetime}
\phantom{blah}

\noindent We now come to perhaps the most unexpected aspect of the formalism we have described - its connection with the so called emergent gravity paradigm. The mathematical results derived here seem to strongly support this paradigm, in which gravitational dynamics is described in terms of thermodynamics of future causal horizon of an event $p_0$. Two of the key ideas in this context -- the use of {\it local Rindler frames} as probes of spacetime curvature (due to Jacobson), and a variational principle based on {\it entropy functional} (due to Padmanabhan et. al.) -- find a unified and purely geometric description in our framework in terms of equi-geodesic {\it surfaces} (which replace the Rindler {\it trajectories}) and the $\lp=0$ term of the coincidence limit of Ricci bi-scalar of the qmetric, which happens to have the same form as the entropy functional. 
\\
\\
\noindent Our formalism amswers one of the most frequently asked questions about the emergent gravity paradigm: \textit{Why choose it over the conventional general relativity based on the Einstein-Hilbert lagrangian}? Most of our conventional intuition about quantum gravity is built on the idea that classical gravity can be obtained as a ``Taylor series expansion" in $\lp^2$ starting from quantum gravity. This, in turn, assumes that all quantum gravitational effects are analytic in $\lp^2$ and will lead to some sensible classical limits when
$\lp \to 0$ limit is taken. Some thought shows that this is a highly questionable assumption and we should take seriously the possibility that quantum gravity
could have features which are non-analytic in $\lp$. In that case, the process of taking limits might involve manipulating singular quantities leading to unexpected
(but interesting) results. Our results imply that this is indeed what happens in the case of gravity, and hence
\\
\\
\textit{Emergent nature of gravitational dynamics, gravitational action, and space and time itself, might be an unmistakable relic of a quantum spacetime endowed with a zero point length.}

	\begin{figure}[H]
	\begin{center}
	\scalebox{0.4}{\includegraphics{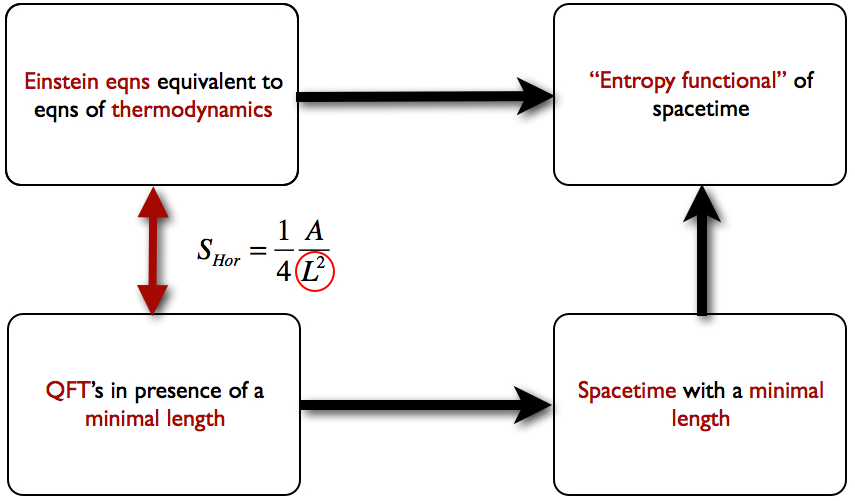}}
	\end{center}
	    \caption{Implications of minimal length for Emergent gravity and spacetime thermodynamics.}%
	    \label{fig:ml-eg}%
	\end{figure}

\noindent A more comprehensive discussion of these ideas developing the emergent gravity paradigm based on above tools and results can be found in the recent reviews by T. Padmanabhan \cite{paddy-EG}.

\section{A broader outlook
\\
\textcolor{white}{bl} -- \textit{bi-tensors} as tools to probe classical and quantum spacetime }
\phantom{blah}

\noindent If one looks deeper into the structure of our fundamental theories -- General Relativity and Quantum Field Theory -- some of the most important mathematical features of these theories, such as geodesic deviation, focussing of geodesics, causal structure, singularity structure of two point functions $G(x,y)$ etc, are characterised in terms of bi-tensors. Characterising the small scale topology of spacetime is yet another aspect which would require a description directly in terms of the distance function $d(p,p_0)$. This contribution describes some very generic aspects of the formalism based on these ideas which were first presented in Ref.~\cite{dk-ml} and developed in further conceptual and technical details in Refs.~\cite{dk-tp-ml, dk-tp-ghy}, and  improved rigor basis in Ref.~\cite{jaffino-dk}. Although our focus has been to reconstruct spacetime metric as a non-local bi-tensor, the basic ideas go beyond this and can be applied to the matter sector as well.
\footnote{The idea of gravity being described fundamentally by a non-local action, with geodesic distance playing the key role, seems to be {\it conceptually} in tune with certain ideas presented in DeWitt and Alvarez et. al.
}
\\
\\
\noindent 
The key underlying idea -- that curvature of spacetime might be solely expressible in terms of behaviour of it's geodesics and related bi-tensorial quantites -- is powerful and more useful from an operational point of view. In absence of non-locality, such a description would eventually coincide with the standard one in terms of local tensors such as $g_{ab}(p), R_{abcd}(p)$ etc. However, if quantum gravity is inherently non-local, then the non-local character of bi-tensors, combined with the non-analytic deformation of geometry necessitated by the scale of non-locality (here brought in through a minimal length), might lead to a very different description of spacetime curvature at smallest of scales; in particular, it may leave a relic independent of the details or value of the short distance cut-off, thereby acting as a crucial guidepost towards our understanding of classical gravity itself.

	\begin{figure}[H]
	\scalebox{0.25}{\includegraphics{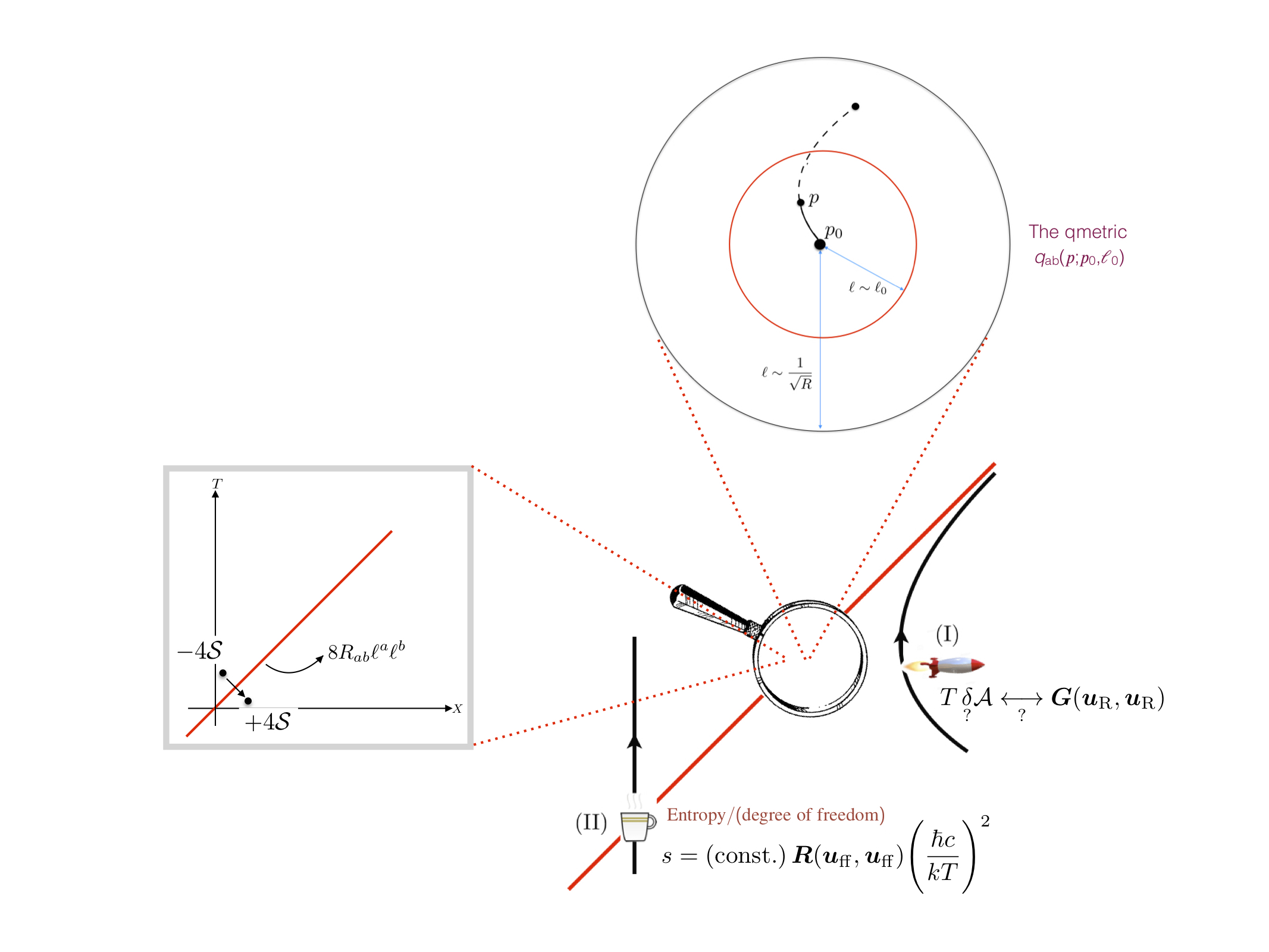}}
	   \caption{Small scale structure of spacetime: Implications and Future outlook}%
	   \label{fig:outlook}%
	\end{figure}

\noindent Since all the relevant comments, remarks and discussions have already been given in the respective sections, I conclude this contribution with a succinct, pictorial summary of the key features of the formalism and its inter-connectedness and relevance for many other important issues in semi-classical and quantum gravity. 
\footnote{
It is my hope that the serious reader who has read through this contribution and looked at the {\it References} given below, would understand what Fig. \ref{fig:outlook} represents!
}
\section*{Acknowledgments}

I thank the organisers of the Tenth International Workshop DICE2022, held at Castiglioncello, for kindly inviting me to present this work at the conference, and Prof Elze in particular for coordinating the entire effort and ensuring it runs smoothly. I would also like to thank T. Padmanabhan, Sumanta Chakraborty, and A. Pesci for discussions and collaboration on several aspects of this work, and Raghvendra Singh for comments on the draft. 

$$$$



\end{document}